\documentstyle[aps,epsf,epsfig,amsmath,amssymb]{revtex}

\newcommand{\be}{\begin{eqnarray}}
\newcommand{\ee}{\end{eqnarray}}
\newcommand{\bea}{\begin{eqnarray}}
\newcommand{\eea}{\end{eqnarray}}
\newcommand{\ba}{\begin{array}}
\newcommand{\ea}{\end{array}}

\newcommand{\nn}{\nonumber \\}

\def\w{\omega}

\def\CM{{\cal M}}

\def\CR{{\cal R}}

\newcommand{\bref}[1]{(\ref{#1})}
\newcommand{\sac}{\, , \qquad}
\newcommand{\bbR}{{\mathbb{R}}}
\newcommand{\bbZ}{{\mathbb{Z}}}

\newcommand{\bbC}{{\mathbb{C}}}
\newcommand{\bbP}{{\mathbb{P}}}

\begin{document}


\twocolumn[\hsize\textwidth\columnwidth\hsize\csname
@twocolumnfalse\endcsname


\title{Supersymmetric $AdS_3$ solutions of type IIB supergravity}
\author{Jerome P. Gauntlett, Ois\'{\i}n A. P. Mac Conamhna, Toni
   Mateos, and Daniel Waldram  \\}
\address{Theoretical Physics Group, Blackett Laboratory,
   Imperial College, London SW7 2AZ, U.K.\\
   The Institute for Mathematical Sciences, Imperial College,
   London, SW7 2PG, U.K.\\}

\maketitle

\begin{abstract}
For every positively curved K\"ahler-Einstein manifold in four
dimensions we construct an infinite family of supersymmetric solutions
of type IIB supergravity. The solutions are warped products of $AdS_3$
with a compact seven-dimensional manifold and have non-vanishing
five-form flux. Via the AdS/CFT correspondence, the solutions are dual
to two-dimensional conformal field theories with $(0,2)$
supersymmetry. The corresponding central charges are rational numbers.

\center{\it Dedicated to Rafael Sorkin in celebration of his 60th
  birthday.}
\end{abstract}

\vskip1pc]

%
%

The AdS/CFT correspondence~\cite{Maldacena:1997re} states that any solution
of string or M-theory with an $AdS_{d+1}$ factor should be equivalent
to a conformal field theory (CFT) in $d$ spacetime dimensions. This
correspondence, and its generalisations, has provided profound insight
into the non-perturbative structure of string theory, the structure of
quantum field theory and the quantum properties of black holes.

Backgrounds with $AdS_3$ factors are of particular interest because,
unlike in higher dimensions, the conformal group in two-dimensions is
infinite dimensional. As a consequence two-dimensional conformal field
theories are much more tractable than their higher dimensional
cousins;
for instance, many models are exactly solvable,
and there is a considerable literature on the subject.
It would be a significant development if, via the AdS/CFT
correspondence, string or M-theory can make contact with
this large body of work.

However, until now there were only a few known explicit $AdS_3\times
\CM$ solutions, with compact $\CM$. The most well studied class of
examples are the $AdS_3\times S^3\times X$ backgrounds of type IIB
supergravity, where $X=T^4$ or $K3$. These are dual to ${\cal
  N}=(4,4)$ conformal field theories that are deformations of the
sigma model based on the orbifold $Sym(X)^n/S_n$. From a string theory
perspective, these backgrounds describe the backreaction of a D-brane
configuration that can be related to a black hole in five
dimensions. It is a remarkable fact that the entropy of this black hole
can be precisely derived from the central charge of the dual conformal
field theory~\cite{Strominger:1996sh}.

There have also been recent investigations
into the conformal field
theory dual to the $AdS_3\times S^3\times S^3\times S^1$ background of
type II string theory \cite{Gukov:2004ym}
(see \cite{CT}--\cite{deBoer:1999rh} for earlier discussions).
Despite the fact that the field theory has a larger version of ${\cal
  N}=(4,4)$ superconformal 
symmetry than those dual to the $AdS_3\times S^3\times X$ solutions,
it has proved more difficult to identify them as a number of
subtleties arise.

The purpose of this letter is to present a new infinite class of
supersymmetric $AdS_3$ backgrounds of type IIB string theory, which
are dual to two-dimensional conformal field theories with ${\cal
  N}=(0,2)$ supersymmetry. It will be very interesting if these
conformal field theories can be explicitly identified. It will also be
very interesting to know whether or not our solutions can be related
to black holes.

The new solutions are warped products of $AdS_3$ with a compact
seven-dimensional manifold $\CM_7$ and have non-trivial self-dual
five-form. The manifold $\CM_7$ is constructed as a $U(1)$ fibration
over a six-dimensional manifold $B_6$. In turn $B_6$ is an $S^2$
bundle over an arbitrary K\"ahler-Einstein manifold $KE_4$ with
positive curvature. Such $KE_4$ manifolds are either $S^2\times S^2$,
$\bbC\bbP^2$ or a del-Pezzo surface $dP_k$ with $k=3,\dots,8$. For
each such $KE_4$ we have an infinite discrete number of explicit
solutions parametrised by two positive integers $p$ and $q$, together
with an integer $n$ which specifies the D3-brane charge.
The fibration structure implies that the group of symmetries
preserving the solutions contains at least two $U(1)$ factors, one of
which corresponds to the R-symmetry of the dual conformal field theory.
The construction is remarkably similar to the construction of
seven-dimensional Sasaki-Einstein manifolds presented in
\cite{Gauntlett:2004hh}, but we do not know of any direct connection.

As we shall show, the standard supergravity computation gives a
rational central charge $c$ for the dual two-dimensional superconformal
field theories. Specifically
\be
\label{central-charge}
c = \frac{9p\,q^2(p+mq)}{3p^2+3mpq+m^2q^2} \, \frac{Mq}{m^2h^2} \,n^2 \,,
\ee
where the integers $m$ and $M$ depend on the specific choice of $KE_4$:
for $S^2\times S^2$ we have $m=2$, $M=8$; for $\bbC\bbP^2$ we have $m=3$, $M=9$;
for the del-Pezzos $dP_k$, we have $m=1$, $M=9-k$. Finally,
$h=\text{hcf}\{M/m^2,q\}$.

The type IIB solutions presented here were constructed from a
much richer set of solutions of $D=11$ supergravity that will be
described elsewhere~\cite{toap}. The latter solutions are warped
products of $AdS_3$ with eight-dimensional manifolds that are
topologically $S^2$ bundles over six-dimensional K\"ahler spaces. In
the special case that the six-dimensional manifold is $KE_4\times
T^2$, dimensional reduction along one leg of the $T^2$ and
$T$-duality along the other leg leads to the solutions presented
here. In the companion paper \cite{toap} we will also show that when
the six-dimensional manifold is $S^2\times S^2\times T^2$, with the
$S^2$ having different radii, we obtain additional generalisations
of the type IIB solutions presented here. The construction of these
new $AdS_3$ solutions has many similarities with the construction of
the $AdS_5$ solutions constructed in \cite{GMSW,GMSW2}. Indeed the
latter references provided key inspiration for the work presented
here and in \cite{toap}.

In the remainder of this letter we present the detailed local form of
the new solutions and then determine the conditions that need to be
imposed in order for the local solutions to extend to global
solutions.

%
%

\vskip .4 cm
\noindent {\bf The local solutions.}
The type IIB solutions have a metric that is a warped product of
$AdS_3$ with a seven dimensional manifold ${\cal M}_7$:
\be\label{metric1}
ds^2&=& L^2 w\left[ ds^2({AdS_3}) + ds^2({\cal M}_7)\right].
\ee
The warp factor, $w$, just depends on the coordinates on ${\cal M}_7$
and hence this metric has all of the isometries of $ds^2(AdS_3)$.
If $KE_4$ is an arbitrary positively curved K\"ahler-Einstein manifold
with metric $ds^2_{KE_4}$ and K\"ahler-form $J$, then the metric on
${\cal M}_7$ is given by
\bea\label{metric2}
ds^2({\cal M}_7)&=&{3 \over 8y} ds^2_{KE_4}
 + {9 dy^2\over 4q(y)}
 + {q(y) D\psi^2\over 16 y^2(y^2-2y+a)}\nn
&&+ {y^2-2y+a \over 4 y^2} Dz^2 \,,
\eea
where $D\psi=d\psi+P$, $Dz=dz-g(y)D\psi$, and
\bea
g(y)&=&  {a-y \over 2(y^2-2y+a)}\,,
\nn
q(y) &=& 4 y^3-9y^2+6a y-a^2 \,.
\eea
Here $a$ is a constant, $dP=J$ and in these coordinates the warp factor is
simply $w=y$. We have chosen normalisations so that $ds^2_{AdS_3}$ is
the metric on a unit radius $AdS_3$ and $\CR=J$, where $\CR$ is the
Ricci form of $KE_4$, and the constant $L$ is arbitrary, reflecting the
scaling symmetry of the type IIB supergravity action. The only other
non-zero type IIB field in the solution, other than the string
coupling $g_s$, is the self-dual five-form which can be written:
\bea\label{5-form}
g_s F_5 = L^4 \,\, \left[ \mbox{vol}_{AdS_3}\wedge \w_2 + J \wedge
   \w_3 \right]\,,
\eea
where
\bea
\label{F-parts}
\w_2&=&-\frac{a}{4}J+\frac{y(a-y)}{2(y^2-2y+a)}dy\wedge D\psi+y \,dy\wedge Dz\,, \nn
\w_3&=&{3(y-a) \over 64y} J \wedge Dz+ {3 a \over 64 y^2} dy \wedge D\psi\wedge Dz \nn
&&+ {3 q(y) \over 128 y(y^2-2y+a)} J\wedge D\psi \,,
\eea
and $\mbox{vol}_{AdS_3}$ is the volume-form of $ds^2(AdS_3)$.
Note both $\partial_\psi$ and $\partial_z$ are Killing vectors, and
thus the symmetry group of the background, including  $F_5$, is at
least $G\times U(1)^2$ where $G$ is the group of the isometries
of $KE_4$ that preserve $J$.

In~\cite{toap} we show how to derive this class of solutions from a
more general family of solutions of $D=11$ supergravity. We also
explicitly discuss the preservation of supersymmetry arguing that the
solutions must be dual to conformal field theories with ${\cal
  N}=(0,2)$ supersymmetry. Furthermore, the form of the Killing
spinors implies that $\partial_\psi$ generates the isometry dual to the
$U(1)_R$ symmetry of the field theory. Here, instead, we show
that we do indeed have a solution by simply comparing with the
elegant analysis of the most general type IIB
supergravity solutions with $AdS_3$ factors and non-vanishing
$F_5$ presented in~\cite{nakwoo}. There it was shown that the
metric $ds^2(\CM_7)$  can always, locally, be
written as a $U(1)$ fibration over a six-dimensional K\"ahler manifold
satisfying some additional properties.
Introducing the new coordinates $\psi=\psi'-2z'$, $z=-2z'$, and
identifying $z'$ as the coordinate on the $U(1)$ fibration, one can
check that our solution satisfies all of the conditions
in~\cite{nakwoo} (one needs to take into account a rescaling of the
five-form flux and also a typo in~(3.22) of~\cite{nakwoo}).

%
%

\vskip .4 cm
\noindent
{\bf Global Analysis.}
We now need to show that the local solution given above can be
defined globally. First we need to fix the global structure of $\CM_7$. We
will assume that $\CM_7$ is an $S^1$ bundle (with the fibre
parametrised by $z$) over a compact six-dimensional base manifold,
$B_6$. The metric on $B_6$ is given by
\bea\label{base}
ds^2({B}_6)={3 \over 8y} ds^2_{KE_4}
 + {9 dy^2\over 4q(y)}
 + {q(y) D\psi^2\over 16 y^2(y^2-2y+a)} \,.
\eea
For a suitable choice of the range of $a$ and $y$, one can take $B_6$
to be an $S^2$ bundle (with the fibre parametrised by $y,\psi$) over
$KE_4$. More precisely, if ${\cal L}$ is the canonical line bundle of
$KE_4$, the $S^2$ bundle is obtained by adding a point to each
fibre. Topologically $\CM_7$ is the same manifold that was used in the
construction of seven-dimensional Sasaki-Einstein metrics found
in~\cite{Gauntlett:2004hh}.

We first need to show that the metric~\bref{base} on $B_6$ is complete and
regular. It has potentially singular points at the roots of the cubic
polynomial $q(y)$, at the roots of the quadratic polynomial
$y^2-2y+a$ and at $y=0$. If we assume that $a\in(0,1)$ then the three
roots $y_i$ of $q(y)$ are real and strictly positive. If we let
$y_1<y_2<y_3$ then $y_1,y_2 \in (0,1)$. Furthermore, $y^2-2y+a$ is
strictly positive in the interval $(y_1,y_2)$. Thus, by choosing the
range of $y \in(y_1,y_2)$ we are left with potential problems only at
$y_1,y_2$, where $g_{yy}$ diverges and $g_{\psi\psi}$
vanishes. However, these are merely coordinate singularities analogous
to those of polar coordinates at the origin of $\bbR$$^2$. Near $y_1$
and $y_2$ (and, in fact, also $y_3$) the $(y,\psi)$ part of the metric
takes the approximate form
\bea \label{polar}
{9 \over 4 q'(y_i)} \left[ dr^2 + {q'(y_i)^2 \over 144 y_i^2
     (y_i^2-2y_i+a) } \, r^2 d\psi^2 \right] \,,
\eea
where we defined $r_i=2 \sqrt{y-y_i}$. The observation that $q'(y)^2-
144 y^2 (y^2-2y+a)=-36 q(y)$ for any $y$, shows that~\bref{polar} is
free from conical singularities if the period of $\psi$ is chosen to
be $\Delta\psi = 2\pi$. Thus the local metric $\bref{base}$ is regular
everywhere in $B_6$ if we restrict $a\in (0,1)$ and
$y\in(y_1,y_2)$. These choices also ensure that the warp factor $w=y$
does not vanish or diverge which would have led to singularities in
the full ten-dimensional solution.

We now turn to showing that the full metric~\bref{metric2} is
consistent with $\CM_7$ being a $U(1)$-bundle over $B_6$. Observe
first that the norm of the Killing vector $\partial _z$ never
vanishes (or diverges) and so the size of the $S^1$ fibre is always
finite. Let us write $Dz=dz-A$ and denote the period of $z$ by $\Delta
z=2\pi l$. For the metric to be well-defined the rescaled one-form $l^{-1}A$
must be a connection on a {\it bona fide} $U(1)$-fibration. This is
equivalent to the condition that the corresponding first Chern class
$\frac{1}{2\pi}l^{-1}dA$ lies in the integer
cohomology\footnote{Generically the first Chern class may include
  torsion elements in $H^2(B_6,\bbZ)$, but as discussed
  in~\cite{Gauntlett:2004hh}, here  $\pi_1(B_6)=0$ so there is no
  torsion and $H^2(B_6,\bbZ)\cong H^2_\text{de Rham}(B_6,\bbZ)$.}
$H^2_\text{de Rham}(B_6,\bbZ)$. We observe first that
\bea
\label{dA1}
dA&=& g(y) J \, + \, g'(y)dy \wedge D\psi,
\eea
is a globally defined 2-form on $B_6$:
the first term is a smooth polynomial times the globally defined
K\"ahler form, and the second is a smooth polynomial times $dy \wedge
D\psi$. The latter 2-form could only be singular at the roots
$y_1,y_2$, but near those points it is approximately $r \, dr \wedge
d\psi$, which is the volume form on $\bbR^2$ in polar coordinates.

The condition that the first Chern class is in $H^2_\text{de
  Rham}(B_6,\bbZ)$ is equivalent to requiring that the corresponding
periods are integral, that is
\bea \label{chern}
P(C) = {1 \over 2\pi} \int_{C} l^{-1} dA_1 \, \in \, \bbZ \,,
\eea
for any curve $C\in H_2(B_6,\bbZ)$. To check this we need a basis of
the free part of $H_2(B_6,\bbZ)$. In fact, such a basis is described
in~\cite{Gauntlett:2004hh} in a very similar setting. Let
$\{\Sigma_a\}$ be a basis for the free part of $H_2(KE_4,\bbZ)$. Then
$\{C_0,C_a\}$ form a basis of the free part of $H_2(B_6,\bbZ)$, where
we take $C_0$ to be the fibre $S^2$ at a fixed point in the $KE_4$
base-space, and $\{C_a\}$ to be the two-cycles $\{\Sigma_a\}$ sitting
at one of the poles of the $S^2$, say $y=y_1$. We find
that\footnote{One might also consider the periods over two cycles at
  the other pole $\{\tilde{C}_a\}$, however these are not independent,
  since the $S^2$ fibration is such that as homology classes
  $\tilde{C}_a=C_a+mn_a C_0$.}
\bea
   P(C_0) &=& l^{-1}\left[g(y_2)-g(y_1)\right] \,, \nn
   P(C_a) &=& l^{-1}g(y_1) m n_a \,,
\eea
where $m$ and $n_a$ are integers related to ${\cal L}$, the canonical
line bundle of $KE_4$. Specifically $m$ is the largest positive
integer $m$ (known as the Fano index) for which there is a line bundle
${\cal N}$ such that ${\cal L}={\cal N}^m$, and $n_a$ are the periods
$n_a=\int_{\Sigma_a}c_1({\cal N})$ of the Chern class of ${\cal N}$. By
construction the $n_a$ are coprime. (In what follows it is useful to
also write the homology class $\Sigma_{\cal N}$, the Poincar\'e dual of
$c_1({\cal N})$, as $\Sigma_{\cal_N}=s^a\Sigma_a$. Again by definition
the $s^a$ are coprime.) It is then easy to see that the periods $P(C)$
are integer if and only if
\bea
g(y_2)-g(y_1)=lq \sac g(y_1)=lp/m \,,
\eea
for some integers $p,q\in\bbZ$. If $p$ and $q$ are relatively prime,
the periods have no common factors and $\CM_7$ is simply
connected. Note that in general $g(y_2)/g(y_1)=(p+mq)/p$ is rational.

To analyse these conditions it is convenient to introduce a different
variable, $x=(4y-a)/3a$, in terms of which the cubic polynomial reads
\bea
\label{q-of-x}
q(y(x))= {a^2 \over 16} \left[ a(1+3x)^3-(1-9x)^2 \right] \,.
\eea
Some algebra then leads to
\be\label{first-root}
\sqrt{x(y_1)} = \left[1+2\tfrac{g(y_2)}{g(y_1)}\right]^{-1} \,.
\ee
It can be checked that for any $a\in(0,1)$, the location of the first
root is such that $x(y_1)<1/9$, and so the condition \bref{first-root}
implies that we can take $p,q>0$. Using \bref{q-of-x} we find that the
parameter $a$ must be rational and of the form
\be \label{a}
a = \frac{m^2q^2(3p+mq)^2(3p+2mq)^2}{4(3p^2+3mpq+m^2q^2)^3} \,,
\ee
and
\be \label{l}
l = \frac{2(3p^2+3mpq+m^2q^2)m}{3p(p+mq)(2p+mq)} \,.
\ee

To summarise, we have shown that for each pair of integers
$p,q$ with $p,q>0$, the background~\bref{metric1},~\bref{metric2}
and~\bref{5-form} gives a regular supersymmetric type IIB supergravity
solution with compact $\CM_7$ provided that the parameter $a$ and the
parameter fixing the period of the $z$-circle are given by~\bref{a}
and~\bref{l}, respectively. If $p$ and $q$ are coprime, $\CM_7$ is
simply-connected.

%
%

\vskip .4 cm
\noindent {\bf Flux quantisation and central charge.}
To ensure that we have a good solution of string theory we need to
show that the five-form flux is globally defined on $\CM_7$ and
furthermore is quantised. The appropriate quantisation condition is
that the periods of $F_5$ are integers:
\be
N(D) = {1 \over (2\pi l_s)^4 } \int_{D} F_5 \, \, \in \,\, \bbZ \,,
\ee
for any five-cycle $D\in H_5(\CM_7,\bbZ)$, where $l_s$ is the string
length.

Recall that we already argued that $dy\wedge D\psi$ is globally
defined, as is $dy$, since it is proportional to $rdr$ at the roots
$y_1$ and $y_2$, while by definition $J$ and $Dz$ are global
forms. Given the expression~\bref{F-parts}, we immediately see that
$F_5$ is globally defined. To check that the periods are quantised we
need a basis for the free part of $H_5(\CM,\bbZ)$. Note first that a
basis for the free part of $H_4(B_6,\bbZ)$ is given by a section of
the $S^2$ bundle over $KE_4$, say at $y_1$ or $y_2$ together with the
$S^2$ fibrations over each basis two-cycle $\Sigma_a\in
H_2(KE_4,\bbZ)$. Since the $U(1)$ bundle over $B_6$ is non-trivial,
all non-trivial five-cycles come from the $U(1)$ fibration over a
four-cycle in $B_6$. Let us label these as follows: $D_0$ denotes the
five-cycle arising from the section $y=y_1$, $\tilde{D}_0$ is the
cycle corresponding to $y=y_2$, and $D_a$ the cycle arising from
$\Sigma_a$. Note that these cycles are not independent. From the $S^2$
fibration structure of $B_6$ we have $D_0=\tilde{D}_0+ms^a D_a$,
while, using similar arguments to those in the appendix of~\cite{GMSW2},
the $U(1)$ fibration is such that $q\tilde{D}_0+ps^aD_a=0$. The
periods of $F_5$ are given by
\bea
N(D_0)&=& - \left(\frac{3 L^4}{64 \pi l_s^4 g_s}\right)
   \left(\frac{m}{p(p+mq)}\right) M(p+mq) \,, \nn
N(\tilde{D}_0)&=& - \left(\frac{3 L^4}{64 \pi l_s^4 g_s}\right)
   \left(\frac{m}{p(p+mq)}\right) Mp \,, \nn
N(D_a)&=& \left(\frac{3 L^4}{64 \pi l_s^4 g_s}\right)
   \left(\frac{m^3}{p(p+mq)}\right)  q \, n_a\,,
\ee
where $M$ is a positive integer depending on the choice of $KE_4$,
given by
\begin{equation*}
   M = \int_{KE_4}c_1({\cal L})\wedge c_1({\cal L})
     = {1\over 4\pi^2}\int_{KE_4} \CR\wedge \CR 
     = m^2 s^an_a \,.
\end{equation*}
These expressions reflect the relations between the
five-cycles mentioned above. The flux quantisation condition, which is
a quantisation for the possible $AdS_3$ radii in string units, is thus
\be
\label{n-def}
\frac{3 L^4}{64\pi l_s^4 g_s}=\frac{p(p+mq)}{hm^3}n \,,
\ee
where $n$ is an arbitrary integer, we are assuming that $p$ and $q$
are coprime and recall that $h=\text{hcf}\{M/m^2,q\}$.

Since the solutions have only non-vanishing five-form flux,
it is natural to interpret them as the
near horizon limit of some configuration of wrapped and/or
intersecting D3-branes after taking into account their backreaction.
The minimal value of $n=1$ would then naturally correspond to the
minimal configuration of D3-branes, with higher values of
$n$ corresponding to the backreacted geometries of $n$ coincident
configurations of such D3-branes. From~\bref{n-def} we see that, as is
standard in the AdS/CFT correspondence, for finite $p$ and $q$, one
can have small $g_s$ together with small curvatures only if $n\gg 1$.

\vskip .1 cm

Having established all the conditions for our solutions to be proper
string theory backgrounds, we now calculate the central charge of the
dual conformal field theories.
It is well known \cite{Brown:1986nw} that the central charge $c$ is fixed by
the $AdS_3$ radius $L$ and the Newton constant $G_{(3)}$
of the effective three-dimensional theory obtained by
compactifying type IIB supergravity on $\CM_7$:
\be
c = {3 L \over 2 G_{(3)}} \,.
\ee
In our conventions, the type IIB supergravity lagrangian reads
\be
{1\over (2\pi)^7 g_s^2l_s^8}  \, \sqrt{-\det{g}} \,R + \ldots \,.
\ee
Integrating this term over $\mathcal{M}_7$ gives the effective
$G_{(3)}$ and hence the rational central charges given
in~\bref{central-charge}. Note that the $n$ dependence
of~\bref{central-charge} is consistent with the comment that the
solution describes $n$ copies of a minimal D3-brane configuration, the
$n^2$ degrees of freedom arising from open strings ending on the $n$
branes. 

The solutions with $KE_4^+=CP^2$ or $S^2\times S^2$ have global
symmetries that include a $U(1)\times U(1)$ factor that leaves the
Killing spinors invariant. As a consequence the dual CFTs will have
exactly marginal $\beta$-deformations for which corresponding
supergravity solutions can be constructed using the technique
of~\cite{lm}. The CFTs dual to the solutions with $KE_4^+=dP_k$ with
$k>4$ have exactly marginal deformations corresponding to the deformations 
of the complex structure of $dP_k$. Thus the only potentially isolated
CFTs are those dual to the solutions with $KE_4^+=dP_3$ and $dP_4$.


\medskip
\noindent
We would like to thank Dmitriy Belov, Dario Martelli and James Sparks
for discussions. OC is supported by EPSRC. DW is supported by the
Royal Society through a University Research Fellowship.

%
%


\end{document}